%% file: mybib.tex
\documentclass[a4paper]{article}

\usepackage{INTERSPEECH2021}
\usepackage{tabularx}
\usepackage{multirow}
\usepackage{cite}
\usepackage{bbm}
\usepackage{etoolbox}
\usepackage{siunitx}
\usepackage{booktabs}
\title{Low Resource German ASR with Untranscribed Data Spoken by Non-native Children - INTERSPEECH 2021 Shared Task SPAPL System}
\name{Jinhan Wang$^1$,  Yunzheng Zhu$^1$, Ruchao Fan$^1$, Wei Chu$^2$, and Abeer Alwan$^1$}
\address{
  $^1$Department of Electrical and Computer Engineering, University of California Los Angeles, USA
  $^2$PAII Inc., USA}
\email{\{wang7875, yunzhengzhu19, fanruchao\}@g.ucla.edu, chuwei129@pingan.com.cn, alwan@ee.ucla.edu}
\begin{document}


\maketitle
\begin{abstract}

This paper describes the SPAPL system for the INTERSPEECH 2021 Challenge: Shared Task on Automatic Speech Recognition for Non-Native Children's Speech in German. $\sim$ 5 hours of transcribed data and $\sim$ 60 hours of untranscribed data are provided to develop a German ASR system for children. For the training of the transcribed data, we propose a non-speech state discriminative loss (NSDL) to mitigate the influence of long-duration non-speech segments within speech utterances. In order to explore the use of the untranscribed data, various approaches are implemented and combined together to incrementally improve the system performance. First, bidirectional autoregressive predictive coding (Bi-APC) is used to learn initial parameters for acoustic modelling using the provided untranscribed data. Second, incremental semi-supervised learning is further used to iteratively generate pseudo-transcribed data. Third, different data augmentation schemes are used at different training stages to increase the variability and size of the training data. Finally, a recurrent neural network language model (RNNLM) is used for rescoring. Our system achieves a word error rate (WER) of 39.68\% on the evaluation data, an approximately 12\% relative improvement over the official baseline (45.21\%). 



  

\end{abstract}

\section{Introduction}
Limited data, untranscribed data, non-native speech, and children's speech are the most difficult challenges for Automatic Speech Recognition (ASR). The data for this Challenge (TLT-2021 Shared Task), which is a continuation of the INTERSPEECH 2020 Challenge \cite{gretter2020overview}, pose all these difficulties. 

State-of-the-art deep learning-based ASR systems require sufficient training data to obtain a competitive result \cite{hinton2012deep}. However, there are many low-resource tasks such as children's ASR. The lack of data affects recognition performance dramatically. To deal with this problem, data augmentation, a common and low-cost solution, was shown to be effective in training robust acoustic models and resulting in lower word error rates (WER) \cite{cui2015data,ko2015audio,tuske2014data,9413801,park2019specaugment,jaitly2013vocal,chen2020data}. Speed perturbation\cite{ko2015audio} and SpecAug\cite{park2019specaugment} are two widely-used augmentation techniques for ASR. Volume perturbation and noise augmentation are selectively used according to the quality of the data \cite{chen2020data}. Particularly for child speech, Chen et al. used pitch perturbation to increase the amount of training data \cite{chen2020data}, while Yeung et al. proposed F0-based normalization and data augmentation methods for child ASR to increase the variability and amount of training data \cite{9413801}. Data augmentation methods are also applied to the same INTERSPEECH Challenge for English ASR last year and shown to be effective in dealing with low-resource transcribed data \cite{knill2020non,lo2020ntnu,shahin2020unsw,kathania2020data}.


Untranscribed data is easier to obtain from various sources. Therefore, many studies investigated how to utilize large amounts of untranscribed data to improve low-resource ASR tasks \cite{imseng2014exploiting,thomas2013deep}. Unsupervised pre-training and semi-supervised learning are two effective methods to deal with lack of transcribed data. For unsupervised pre-training methods, the input speech signal itself is regarded as a supervision so that the model can either predict frames using their history dependencies such as autoregressive predictive coding (APC) \cite{chung2019unsupervised,oord2018representation,ravi2020exploring,fan2021bi}, or using BERT pre-training logistics to reconstruct masked input frames from unmasked frames \cite{wang2020unsupervised,jiang2019improving}. By pre-training the up-stream task using untranscribed data, a model can learn general speech representations and transfer the learned knowledge to the down-stream task by a fine-tuning process. 
On the other hand, semi-supervised learning in ASR also yields improvements by leveraging untranscribed data \cite{lamel2002lightly,synnaeve2019end,weninger2020semi,khonglah2020incremental,dey2019exploiting,cui2012multi}. After training on the limited transcribed data, pseudo labels for untranscribed data are generated to further train the system. For conventional one-shot semi-supervised learning, different pseudo label types, at the frame-level and sequence-level, are investigated, and resulting in improved ASR system performance over the supervised system \cite{singh2020large,kahn2020self}.
In \cite{synnaeve2019end, khonglah2020incremental}, compared with one-shot semi-supervised learning, incremental semi-supervised learning consistently prevails, where unsupervised data's pseudo labels are generated iteratively from incrementally updated models. 






In this paper, we develop a hidden markov model-deep neural network (HMM-DNN) hybrid system using a bidirectional long short-term memory (BLSTM) model for the TLT-2021 German ASR Challenge for non-native children's speech and we propose several methods to improve the performance of the system. The reason we choose BLSTM as our baseline is that BLSTM outperforms time delay neural network (TDNN) since BLSTM considers the information from both directions in sequential data \cite{fan2021bi}. The challenge provides 5 hours of transcribed data and 60 hours of untranscribed data. When training an acoustic model with the 5 hours of transcribed data, we observe the dominance of the non-speech segments in each utterance. In order to solve this problem, we propose a non-speech state discriminative loss (NSDL) strategy to prevent overfitting to non-speech states. In order to use the untranscribed data, bidirectional autoregressive predicitve coding (Bi-APC) \cite{fan2021bi} and incremental semi-supervised learning (SSL) \cite{manohar2018semi} are combined together to iteratively improve the system performance. At each training stage, different data augmentation schemes are considered to increase the variability of the training data. We further train a recurrent neural network language model (RNNLM) for rescoring. As a result, the system ranked third for the German closed track and second for the open track. 

The remainder of this paper is organized as follows: Section 2 presents dataset description and the strategies used in our system development along with the baseline settings. Experimental results and discussion are given in Section 3. Section 4 concludes the paper.

\section{Dataset and System Development}


In this section, we first describe the dataset and the official baseline system in the Kaldi toolkit \cite{povey2011kaldi}. We then describe our developed baseline using Pykaldi2 toolkit \cite{lu2019pykaldi2} and the proposed methods for improving the system.

\subsection{Dataset}

The provided speech data \cite{gretter2020tlt} contain $\sim$5 hours of transcribed training data from 296 pupils, about 1445 utterances, and $\sim$1 hour of transcribed development data from 72 pupils, about 339 utterances, and $\sim$60 hours of untranscribed training data from 124 pupils, about 10047 utterances. Two text sources are provided: (1) manual transcriptions for 5 hours of the training data, and (2) written data extracted from the sentences written by the pupils. Participants' ages range from 9 to 16. 
\subsection{Official Baseline}
The official baseline system is obtained based on a Kaldi recipe. 39-dimensional mel-frequency cepstral coefficients (MFCCs), including first and second derivatives are used to train the gaussian mixture model (GMM). The provided multi-lingual lexicon is used to additionally model English and Italian words because the mother tongue of the children is English or Italian. A 3-way speed perturbation strategy is applied to the transcribed and untranscribed datasets with speed factors of 0.9, 1, and 1.1. The acoustic model is based on TDNNs. The inputs to the TDNNs comprise 40-dimensional high-resolution MFCCs, without using derivatives. A TDNN model is first trained with the 3-fold transcribed data, and then fine-tuned with a weighted combination of 3-fold transcribed and 3-fold untranscribed data using one-shot SSL. A 4-gram language model is trained using the transcriptions of the training set.




\subsection{Developed BLSTM Baseline}
For the convenience of implementing unsupervised pre-training and semi-supervised training algorithms, we use Pykaldi2, a PyTorch-based toolkit, for system development. 80-dim log-mel-filter bank features are first extracted every 10ms with a 25ms Hamming window. An additional frame of features after each frame is then appended to form a 160-dim input. The model consists of 4 BLSTM blocks with 512 hidden units in each direction, followed by a linear layer that transforms the output of BLSTMs to HMM state spaces. In each BLSTM block, batch normalization is used. The number of output units is 2392 obtained from clustering all tri-phone states. The dropout rate in the network is empirically set to 0.3. 

Before training the baseline, we observed that there are some non-speech utterances in the provided transcribed dataset. Hence, 
211 non-speech utterances out of 1445 utterances are eliminated in the transcribed training data.

The model is first trained with a sequence-wise mechanism where each utterance is fed into the model as one instance (1169 frames on average). However, as shown in Table \ref{tab:duration_table}, the variance of the duration in utterances are large, leading to many unnecessary paddings. Moreover, BLSTM does not model sentences over 3,000 frames well. To mitigate the gradient vanishing problem in BLSTMs for modelling long-duration utterances and accelerate the training, the model is then trained in chunk-wise mechanism where one instance is a segment consist of 300 frames with 10 appended neighboring frames. The appended frames are only used in training for accumulating the BLSTM hidden states.

\subsection{Data Augmentation Schemes}

To mitigate the low resource issue for the dataset and to increase robustness of the model, data augmentation methods are used in different stages of the training recipe. Pitch perturbation, volume perturbation, speed perturbation, noise augmentation, and vocal tract length perturbation (VTLP) are considered for the transcribed data. Speed perturbation and noise augmentation are used for the untranscribed dataset. For noise augmentation, the removed non-speech utterances are used as foreground and background noise to augment the original audio with various random SNRs. Speed perturbation uses a conventional method with 3-way warping factor of 0.9, 1, and 1.1. Volume perturbation, pitch perturbation, and VTLP are implemented using Kaldi scripts with random warping factors. We also tried SpecAug for our experiments, but the technique does not improve performance. We will detail the data augmentation schemes for each training stage in Section 3. 

\subsection{Non-speech State Discriminative Loss (NSDL)}
The provided dataset in this Challenge contains many long-duration non-speech segments within speech utterances. The statistic of non-speech and speech states, obtained using forced alignment from a GMM model, are shown in Table~\ref{tab:duration_table}. The table shows that the number of non-speech states is more than twice the number of the speech states, which may result in more deletion errors because of overfitting to the non-speech states. 
In order to balance the data for acoustic model training and to effectively discriminate between non-speech and speech states, we decompose the probability distribution over all pdf-ids into two parts and propose a NSDL function ($L_\text{NSDL}$), replacing cross entropy loss during training. 



\begin{table}[t]
  \caption{Statistic of non-speech and speech states in the 5-hour transcribed data.}
  \label{tab:duration_table}
  \centering
  \begin{tabularx}{190pt}{c c c c}
    \toprule
    \multicolumn{1}{c}{\textbf{HMM States}} &  \multicolumn{1}{c}{\textbf{Total}} &
    \multicolumn{1}{c}{\textbf{Avg (Utt)}} &
    \multicolumn{1}{c}{\textbf{Std (Utt)}}
    \\
    \midrule
    Non-speech & 11x10$^5$ & 817 &    875             \\
    Speech & 5x10$^5$ & 352  &   519             \\

    \bottomrule
  \end{tabularx}
\end{table}

\begin{figure}[t]
  \centering
  \includegraphics[width=8cm,height=7.5cm]{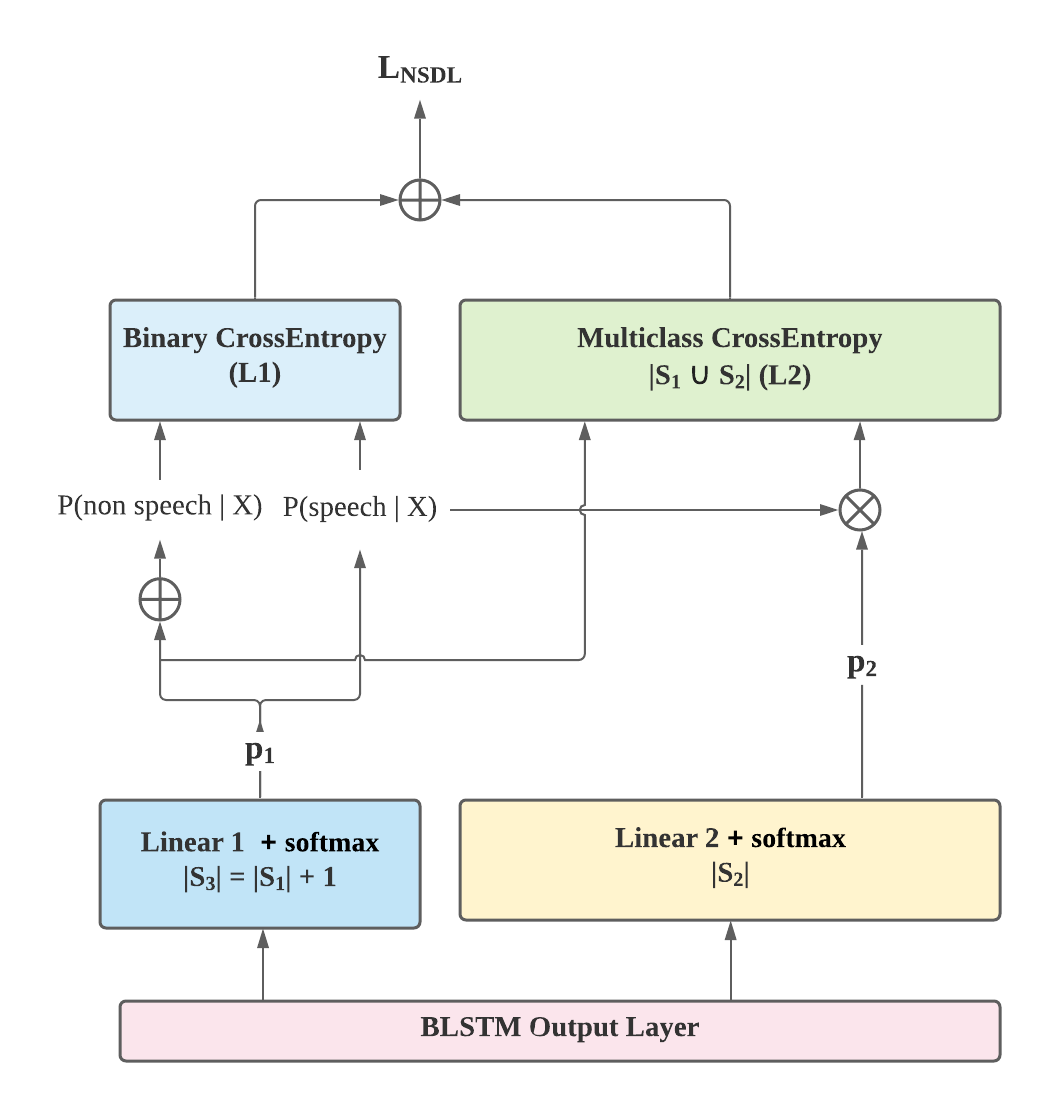}
  \caption{A Schematic Diagram of NSDL}
  \label{fig:SDL}
\end{figure}




The schematic diagram of NSDL is shown in Figure~\ref{fig:SDL}. Suppose the sets of non-speech states and speech states are $S_1$ and $S_2$, respectively. The original acoustic model outputs a probability distribution over all states in $S_1\cup S_2$. Instead of doing a single classification task among all HMM states, the last BLSTM layer output is fed into two separate fully connected layers to obtain two probability distributions $p_1$ and $p_2$ over states in $S_3=\{S_1, speech\}$ and $S_2$, respectively. The symbol \emph{speech} is a placeholder reserved to represent a general speech state, regardless of specific phoneme states. The probability of a state belong to \emph{speech} state is denoted as $P_{\text{speech}}$. We can also obtain $P_{\text{non-speech}}$ by summing up all non-speech states in $S_1$. The two probabilities are used to construct a speech/non-speech binary classifier, and is shown in the left path of Figure~\ref{fig:SDL}. The output probability distributions $p_1$ and $p_2$ are combined to construct the original HMM states classification problem as plotted in the right path of Figure~\ref{fig:SDL}. Let $X$ be the input sequence and $Y=\{y_1,...,y_t,...,y_T\}$ represents the results from forced alignment, the ground truth for binary classification is $b_t=\mathbbm{1}(y_t\in S_2)$. The original acoustic model training is reformulated into a multi-task training as:

\begin{flalign}
    &P(\text{non-speech}|X) = \sum_{s\in S_1}p_1(s|X) \\
&\begin{aligned}
    L_1 = -\sum_{t=1}^T(&\mathbbm{1}(b_t==1)\log(p_1(\text{speech}|X)) + \\
    &\mathbbm{1}(b_t==0)\log(P(\text{non-speech}|X)))
\end{aligned} \\
    &P(s|X) = 
    \begin{cases}
    p_1(s | X) & s\in S_1 \\
    p_1(\text{speech}|X)p_2(s|X) & s\in S_2
    \end{cases} \\
    &L_2 = -\sum_{t=1}^T\log(P(s = y_t|X)) \qquad \\
    &L_{\text{NSDL}} = L_1 + \lambda{L_2}
\end{flalign}
where \textit{$L_1$} is the loss function for the binary classifier for non-speech and speech states, and \textit{$L_2$} is the cross entropy loss for the classifier among all HMM states. $\lambda$ denotes the task ratio for \textit{$L_2$}, which is empirically set to 1. 


In addition, we apply weights to the loss with respect to non-speech/speech classes in $L_2$ to further alleviate the problem that the model is more likely to classify frames as non-speech. The weights of non-speech states and speech states are set to be 0.9 and 1. Hence, the model will assign a larger penalty when it mis-classifies a speech frame as a non-speech frame.


\subsection{Bidirectional Autoregressive Predictive Coding}
Since the provided dataset contains around 60 hours of untranscribed data, we apply Bi-APC \cite{fan2021bi} as an unsupervised pre-training strategy to learn common knowledge from this dataset and then transfer the learned knowledge into the down-stream 5-hour low-resource ASR task. The core mechanism of Bi-APC is to predict a frame \textit n steps after the current frame given all past frames, and predict a frame \textit n steps before the current frame given all future frames in a bi-directional manner. In Bi-APC, BLSTM is split into forward and backward paths, such that each path cannot be supervised in reverse order to prevent equivalent mapping. The model is pre-trained by optimizing the sum of the mean absolute error from both directions. The \textit n is set to 2 in our experiments. We pre-trained the model for 20 epochs. Due to the page limitation, please refer to \cite{fan2021bi} for more details.


\subsection{Incremental Semi-Supervised Learning (SSL)}
Instead of generating pseudo labels for all untranscribed data in the official baseline, we apply SSL for several iterations and train them incrementally. In each iteration, we decode the entire untranscribed dataset with the best model at that moment and filter out those utterances with low log-likelihood. The goal is to secure the quality of the pseudo transcription. The chosen data are appended to the transcribed data for the next iteration. Log-likelihood thresholds are determined according to the corresponding WER of the development set and the number of available sentences remaining after filtering. We only apply three iterations due to the time limitation and insignificant improvement in WERs of the development and evaluation datasets after the third iteration. 

\subsection{RNN Language Model Rescoring}


All the provided transcriptions and text from written data are used to train a RNNLM \cite{mikolov2011rnnlm,mikolov2012statistical} for second-pass rescoring using the script in Kaldi. The model consists of two LSTM layers with a hidden size of 128. The dimension of the word embedding is 256. During rescoring, we make a grid search with language model weights 0.25, 0.3, 0.35 and n-gram order 2 and 3. The best settings for the development set in each experiment are used to rescore the lattice of the evaluation set.



\section{Experiments and Discussion}
Experiments are conducted using Pykaldi2 for acoustic model training and Kaldi for decoding. The results of the ASR experiments using different methods are shown in Table \ref{tab:res}.

\input{Tables/Table1.tex}
\subsection{Baselines}
We first develop an HMM-DNN hybrid system with BLSTM modelling as our baseline. In addition to sequence-wise model training, we experiment with chunk-wise training using a chunk size of 300 frames, and left and right context chunks of 10 frames each. The table shows that chunk-wise training improves the performance over sequence-wise training mechanism by 5.68\% and 11.55\% in relative WER for the development and evaluation datasets, respectively. The official baseline is also included in the table.

\subsection{NSDL}

When using our proposed NSDL method, we further achieve a relative improvement of 5.07 \% and 3.09 \% on the development and evaluation datasets, respectively, compared with the chunk-wise training mechanism. A more fair baseline for NSDL is using the default VAD in Kaldi to filter out the silence frames, which has a WER of 54.33 \% for the development set. The result is even worse than the model without VAD. The reason may be that many non-speech states are not silence states, but laugh, hesitation and noise. The proposed NSDL method is more suitable in this case. Starting from Section 3.3, we use the untranscribed data to further improve the performance. The relative WER will be in reference to the official baseline.


\subsection{Bi-APC}
The first method we used to train the 60 hours of untranscribed data is Bi-APC.
Using the same data augmentation strategy for transcribed data, we apply 3-way speed perturbation to the untranscribed data. As shown in Table \ref{tab:res}, a better performance is obtained by using Bi-APC. Both WERs for the development and evaluation data are further improved, and relative improvements of 4.72\% and 1.35\%, respectively, are observed, compared to the official baseline system.

\subsection{Incremental SSL}
Incremental SSL and various augmentation strategies are applied. 
As shown in Table \ref{tab:res}, the first iteration of SSL chose 0.35 as the log-likelihood threshold, by applying a 10-fold (2 pitch + 3 volume + 2 VTLP + 3 speed) augmentation on the transcribed data and 3-way speed perturbation on untranscribed data. The WER does not improve significantly. The second iteration with a 0.3 threshold, with the same augmentation strategies used in the first iteration, results in a relative improvement of 6.36\% and 9.07\% for the development and evaluation datasets, respectively, compared to the official baseline. When applying a third iteration with a threshold of 0.28 and the same augmentation strategies, the performance is only improved for the development data with the WER of 47.10\%. Thus, we do not apply another SSL iteration, but implement different augmentation strategies at this stage. When we apply one iteration of SSL to the BLSTM baseline, the performance still improves from 57.54\% to 56.86\% in WER for the development set.

Replacing 2-fold VTLP with 2-fold noise perturbation for the transcribed data and keeping the untranscribed data augmentation to be 3-fold speed perturbation results in a performance degradation. Adding only 2-fold noise perturbation to the untranscribed dataset does not improve the performance. However, 12-fold (2 pitch + 3 volume + 2 VTLP + 2 Noise + 3 speed) augmentation results in better performance for the evaluation dataset, though slightly worse for the development dataset.







\subsection{RNNLM Rescoring}
RNNLM rescoring (in parentheses in Table \ref{tab:res}) is applied to the output of the third iteration of SSL, which results in a relative improvement of approximately 2\% for both the development and evaluation datasets. The best performance is achieved with a 12-fold augmentation, resulting in a WER of 39.68 \%, a 12.23\% relative improvement 
over the official baseline system.
If we use the development set in training, then the WER for evaluation set improves to 38.85\%. 

\section{Summary and Conclusion}
In this paper, we describe the SPAPL ASR system for the INTERSPEECH 2021 Challenge: Shared Task on Automatic Speech Recognition for Non-Native Children's Speech in German (closed track). To compensate for long-duration non-speech segments within speech utterances, we propose a novel non-speech discriminative loss in the acoustic model training phase to enable classification of the speech/non-speech states. To explore the use of untranscribed data, Bi-APC pre-training and incremental semi-supervised learning are combined together to iteratively improve the system performance. A RNN language model is also trained for rescoring. The final system, which incorporates all the methods mentioned above, achieves a 39.68\% WER on the evaluation dataset which is the third place in the closed track. If we include the development dataset in training, the WER becomes 38.85\%. In the future, we plan to apply discriminative training using Maximum Mutual Information (MMI) as an objective function, and further investigate other language models for rescoring. 

\section{Acknowledgements}
This work was supported in part by the NSF. 

\bibliographystyle{IEEEtran}

\bibliography{mybib}


\end{document}

%% file: Tables/Table1.tex
\begin{table*}[th]
  \caption{\% Word Error Rate (WER) for different systems on the TLT-2021 development and evaluation sets, including chunk-wise modelling, NSDL, Bi-APC, different iterations of SSL, and LM rescoring (in parentheses). 
  The second and the third column represents augmentation schemes for the transcribed and untranscribed datasets respectively. SP: speed perturbation, Pit.: pitch perturbation, Vol.: volume perturbation, VTLP: vocal tract length perturbation, and Noise: noise perturbation. The last line (+dev) indicates that the  development data is included in the training for the open track case. Best performance in each case is bold-faced.}
  \label{tab:res}
  \centering
  \begin{tabularx}{453pt}{lcccc}
    \toprule
    \multicolumn{1}{l}{\textbf{Model}} &  \multicolumn{1}{c}{\textbf{Transcribed Data.}} &
    \multicolumn{1}{c}{\textbf{Untranscribed Data}} &
    \multicolumn{1}{c}{\textbf{Dev(\%)}} & \multicolumn{1}{c}{\textbf{Eval(\%)}} \\
    \midrule
    \midrule
    Official Baseline & 3x SP & $-$ & 52.38 & 45.21             \\
    \midrule
    \multicolumn{5}{c}{Proposed Systems} \\
    \midrule
    BLSTM Baseline& 3x SP & $-$ & 57.54 & 56.63               \\
    + chunk & 3x SP  & $-$ & 54.27 & 50.09 \\
    \hspace{2.8mm}+ NSDL & 3x SP  & $-$ & 51.52 & 48.54              \\
    \hspace{5mm}+ Bi-APC & 3x SP  & 3x SP & 49.91 & 44.60               \\
    \midrule
    \hspace{0mm}+ SSL iter1 (0.35) & 2x Pit. 3x Vol. 2x VTLP 3x SP & 3x SP & 49.91 & 43.55             \\
    \hspace{2.5mm}+ SSL iter2 (0.3) & 2x Pit. 3x Vol. 2x VTLP 3x SP & 3x SP & 49.05 & 41.11               \\
    \midrule
    \hspace{5mm}\multirow{4}{*}{+ SSL iter3 (0.28)}
     & 2x Pit. 3x Vol. 2x VTLP 3x SP  & 3x SP & \textbf{47.10(46.30)} & 41.23(41.05)              \\
    
    & 2x Pit. 3x Vol. 2x VTLP 3x SP  & 2x Noise 3x SP & 47.79(46.99) & 41.23(40.81)     \\
    & 2x Pit. 3x Vol. 2x Noise 3x SP & 3x SP & 47.22(46.59) & 41.28(39.86)              \\
    & 2x Pit. 3x Vol. 2x VTLP 2x Noise 3x SP &  3x SP & 48.02(46.61) & \textbf{40.87(39.68)} \\
    
    \midrule
    \hspace{0mm}+ dev& 2x Pit. 3x Vol. 2x VTLP 2x Noise 3x SP &  3x SP & $-$ & 39.92(38.85)               \\
    \bottomrule
  \end{tabularx}
  
\end{table*}

%% file: mybib.bbl
\begin{thebibliography}{10}
\providecommand{\url}[1]{#1}
\csname url@samestyle\endcsname
\providecommand{\newblock}{\relax}
\providecommand{\bibinfo}[2]{#2}
\providecommand{\BIBentrySTDinterwordspacing}{\spaceskip=0pt\relax}
\providecommand{\BIBentryALTinterwordstretchfactor}{4}
\providecommand{\BIBentryALTinterwordspacing}{\spaceskip=\fontdimen2\font plus
\BIBentryALTinterwordstretchfactor\fontdimen3\font minus
  \fontdimen4\font\relax}
\providecommand{\BIBforeignlanguage}[2]{{%
\expandafter\ifx\csname l@#1\endcsname\relax
\typeout{** WARNING: IEEEtran.bst: No hyphenation pattern has been}%
\typeout{** loaded for the language `#1'. Using the pattern for}%
\typeout{** the default language instead.}%
\else
\language=\csname l@#1\endcsname
\fi
#2}}
\providecommand{\BIBdecl}{\relax}
\BIBdecl

\bibitem{gretter2020overview}
R.~Gretter, M.~Matassoni, D.~Falavigna, K.~Evanini, and C.~W. Leong, ``Overview
  of the interspeech \uppercase{TLT}2020 shared task on \uppercase{ASR} for
  non-native children’s speech,'' \emph{Proc. Interspeech 2020}, pp.
  245--249, 2020.

\bibitem{hinton2012deep}
G.~Hinton, L.~Deng, D.~Yu, G.~E. Dahl, A.-r. Mohamed, N.~Jaitly, A.~Senior,
  V.~Vanhoucke, P.~Nguyen, T.~N. Sainath \emph{et~al.}, ``Deep neural networks
  for acoustic modeling in speech recognition: The shared views of four
  research groups,'' \emph{IEEE Signal processing magazine}, vol.~29, no.~6,
  pp. 82--97, 2012.

\bibitem{cui2015data}
X.~Cui, V.~Goel, and B.~Kingsbury, ``Data augmentation for deep neural network
  acoustic modeling,'' \emph{IEEE/ACM Transactions on Audio, Speech, and
  Language Processing}, vol.~23, no.~9, pp. 1469--1477, 2015.

\bibitem{ko2015audio}
T.~Ko, V.~Peddinti, D.~Povey, and S.~Khudanpur, ``Audio augmentation for speech
  recognition,'' in \emph{Sixteenth Annual Conference of the International
  Speech Communication Association}, 2015.

\bibitem{tuske2014data}
Z.~T{\"u}ske, P.~Golik, D.~Nolden, R.~Schl{\"u}ter, and H.~Ney, ``Data
  augmentation, feature combination, and multilingual neural networks to
  improve \uppercase{ASR} and \uppercase{KWS} performance for low-resource
  languages,'' in \emph{Fifteenth Annual Conference of the International Speech
  Communication Association}, 2014.

\bibitem{9413801}
G.~Yeung, R.~Fan, and A.~Alwan, ``Fundamental frequency feature normalization
  and data augmentation for child speech recognition,'' in \emph{ICASSP 2021 -
  2021 IEEE International Conference on Acoustics, Speech and Signal Processing
  (ICASSP)}, 2021, pp. 6993--6997.

\bibitem{park2019specaugment}
D.~S. Park, W.~Chan, Y.~Zhang, C.-C. Chiu, B.~Zoph, E.~D. Cubuk, and Q.~V. Le,
  ``Specaugment: A simple data augmentation method for automatic speech
  recognition,'' \emph{Proc. Interspeech 2019}, pp. 2613--2617, 2019.

\bibitem{jaitly2013vocal}
N.~Jaitly and G.~E. Hinton, ``Vocal tract length perturbation
  (\uppercase{VTLP}) improves speech recognition,'' in \emph{Proc. ICML
  Workshop on Deep Learning for Audio, Speech and Language}, vol. 117, 2013.

\bibitem{chen2020data}
G.~Chen, X.~Na, Y.~Wang, Z.~Yan, J.~Zhang, S.~Ma, and Y.~Wang, ``Data
  augmentation for children's speech recognition--the "\uppercase{Ethiopian}"
  system for the slt 2021 children speech recognition challenge,'' \emph{arXiv
  preprint arXiv:2011.04547}, 2020.

\bibitem{knill2020non}
K.~M. Knill, L.~Wang, Y.~Wang, X.~Wu, and M.~J. Gales, ``Non-native
  children’s automatic speech recognition: the interspeech 2020 shared task
  \uppercase{ALTA} systems,'' \emph{Proc. Interspeech 2020}, pp. 255--259,
  2020.

\bibitem{lo2020ntnu}
T.-H. Lo, F.-A. Chao, S.-Y. Weng, and B.~Chen, ``The \uppercase{NTNU} system at
  the interspeech 2020 non-native children’s speech asr challenge,''
  \emph{Proc. Interspeech 2020}, pp. 250--254, 2020.

\bibitem{shahin2020unsw}
M.~Shahin, R.~Lu, J.~Epps, and B.~Ahmed, ``\uppercase{UNSW} system description
  for the shared task on automatic speech recognition for non-native
  children’s speech,'' \emph{Proc. Interspeech 2020}, pp. 265--268, 2020.

\bibitem{kathania2020data}
H.~Kathania, M.~Singh, T.~Gr{\'o}sz, and M.~Kurimo, ``Data augmentation using
  prosody and false starts to recognize non-native children’s speech,''
  \emph{Proc. Interspeech 2020}, pp. 260--264, 2020.

\bibitem{imseng2014exploiting}
D.~Imseng, B.~Potard, P.~Motlicek, A.~Nanchen, and H.~Bourlard, ``Exploiting
  un-transcribed foreign data for speech recognition in well-resourced
  languages,'' in \emph{2014 IEEE International Conference on Acoustics, Speech
  and Signal Processing (ICASSP)}.\hskip 1em plus 0.5em minus 0.4em\relax IEEE,
  2014, pp. 2322--2326.

\bibitem{thomas2013deep}
S.~Thomas, M.~L. Seltzer, K.~Church, and H.~Hermansky, ``Deep neural network
  features and semi-supervised training for low resource speech recognition,''
  in \emph{2013 IEEE international conference on acoustics, speech and signal
  processing}.\hskip 1em plus 0.5em minus 0.4em\relax IEEE, 2013, pp.
  6704--6708.

\bibitem{chung2019unsupervised}
Y.-A. Chung, W.-N. Hsu, H.~Tang, and J.~Glass, ``An unsupervised autoregressive
  model for speech representation learning,'' \emph{Proc. Interspeech 2019},
  pp. 146--150, 2019.

\bibitem{oord2018representation}
A.~v.~d. Oord, Y.~Li, and O.~Vinyals, ``Representation learning with
  contrastive predictive coding,'' \emph{arXiv preprint arXiv:1807.03748},
  2018.

\bibitem{ravi2020exploring}
V.~Ravi, R.~Fan, A.~Afshan, H.~Lu, and A.~Alwan, ``Exploring the use of an
  unsupervised autoregressive model as a shared encoder for text-dependent
  speaker verification,'' \emph{Proc. Interspeech 2020}, pp. 766--770, 2020.

\bibitem{fan2021bi}
R.~Fan, A.~Afshan, and A.~Alwan, ``Bi-apc: Bidirectional autoregressive
  predictive coding for unsupervised pre-training and its application to
  children’s asr,'' in \emph{ICASSP}.\hskip 1em plus 0.5em minus 0.4em\relax
  IEEE, 2021, pp. 7023--7027.

\bibitem{wang2020unsupervised}
W.~Wang, Q.~Tang, and K.~Livescu, ``Unsupervised pre-training of bidirectional
  speech encoders via masked reconstruction,'' in \emph{ICASSP}.\hskip 1em plus
  0.5em minus 0.4em\relax IEEE, 2020, pp. 6889--6893.

\bibitem{jiang2019improving}
D.~Jiang, X.~Lei, W.~Li, N.~Luo, Y.~Hu, W.~Zou, and X.~Li, ``Improving
  transformer-based speech recognition using unsupervised pre-training,''
  \emph{arXiv preprint arXiv:1910.09932}, 2019.

\bibitem{lamel2002lightly}
L.~Lamel, J.-L. Gauvain, and G.~Adda, ``Lightly supervised and unsupervised
  acoustic model training,'' \emph{Computer Speech \& Language}, vol.~16,
  no.~1, pp. 115--129, 2002.

\bibitem{synnaeve2019end}
G.~Synnaeve, Q.~Xu, J.~Kahn, T.~Likhomanenko, E.~Grave, V.~Pratap, A.~Sriram,
  V.~Liptchinsky, and R.~Collobert, ``End-to-end \uppercase{ASR}: from
  supervised to semi-supervised learning with modern architectures,''
  \emph{arXiv preprint arXiv:1911.08460}, 2019.

\bibitem{weninger2020semi}
F.~Weninger, F.~Mana, R.~Gemello, J.~Andr{\'e}s-Ferrer, and P.~Zhan,
  ``Semi-supervised learning with data augmentation for end-to-end
  \uppercase{ASR},'' \emph{Proc. Interspeech 2020}, pp. 2802--2806, 2020.

\bibitem{khonglah2020incremental}
B.~Khonglah, S.~Madikeri, S.~Dey, H.~Bourlard, P.~Motlicek, and J.~Billa,
  ``Incremental semi-supervised learning for multi-genre speech recognition,''
  in \emph{ICASSP 2020-2020 IEEE International Conference on Acoustics, Speech
  and Signal Processing (ICASSP)}.\hskip 1em plus 0.5em minus 0.4em\relax IEEE,
  2020, pp. 7419--7423.

\bibitem{dey2019exploiting}
S.~Dey, P.~Motlicek, T.~Bui, and F.~Dernoncourt, ``Exploiting semi-supervised
  training through a dropout regularization in end-to-end speech recognition,''
  \emph{Proc. Interspeech 2019}, pp. 734--738, 2019.

\bibitem{cui2012multi}
X.~Cui, J.~Huang, and J.-T. Chien, ``Multi-view and multi-objective
  semi-supervised learning for hmm-based automatic speech recognition,''
  \emph{IEEE Transactions on Audio, Speech, and Language Processing}, vol.~20,
  no.~7, pp. 1923--1935, 2012.

\bibitem{singh2020large}
K.~Singh, V.~Manohar, A.~Xiao, S.~Edunov, R.~Girshick, V.~Liptchinsky,
  C.~Fuegen, Y.~Saraf, G.~Zweig, and A.~Mohamed, ``Large scale weakly and
  semi-supervised learning for low-resource video \uppercase{ASR},''
  \emph{Proc. Interspeech 2020}, pp. 3770--3774, 2020.

\bibitem{kahn2020self}
J.~Kahn, A.~Lee, and A.~Hannun, ``Self-training for end-to-end speech
  recognition,'' in \emph{ICASSP 2020-2020 IEEE International Conference on
  Acoustics, Speech and Signal Processing (ICASSP)}.\hskip 1em plus 0.5em minus
  0.4em\relax IEEE, 2020, pp. 7084--7088.

\bibitem{manohar2018semi}
V.~Manohar, H.~Hadian, D.~Povey, and S.~Khudanpur, ``Semi-supervised training
  of acoustic models using lattice-free mmi,'' in \emph{ICASSP}.\hskip 1em plus
  0.5em minus 0.4em\relax IEEE, 2018, pp. 4844--4848.

\bibitem{povey2011kaldi}
D.~Povey, A.~Ghoshal, G.~Boulianne, L.~Burget, O.~Glembek, N.~Goel,
  M.~Hannemann, P.~Motlicek, Y.~Qian, P.~Schwarz \emph{et~al.}, ``The kaldi
  speech recognition toolkit,'' in \emph{IEEE 2011 workshop on automatic speech
  recognition and understanding}, no. CONF.\hskip 1em plus 0.5em minus
  0.4em\relax IEEE Signal Processing Society, 2011.

\bibitem{lu2019pykaldi2}
L.~Lu, X.~Xiao, Z.~Chen, and Y.~Gong, ``Pykaldi2: Yet another speech toolkit
  based on kaldi and pytorch,'' \emph{arXiv preprint arXiv:1907.05955}, 2019.

\bibitem{gretter2020tlt}
R.~Gretter, M.~Matassoni, S.~Bann{\`o}, and F.~Daniele,
  ``\uppercase{TLT}-school: a corpus of non native children speech,'' in
  \emph{Proceedings of The 12th Language Resources and Evaluation Conference},
  2020, pp. 378--385.

\bibitem{mikolov2011rnnlm}
T.~Mikolov, S.~Kombrink, A.~Deoras, L.~Burget, and J.~Cernocky,
  ``\uppercase{RNNLM}-recurrent neural network language modeling toolkit,'' in
  \emph{Proc. of the 2011 ASRU Workshop}, 2011, pp. 196--201.

\bibitem{mikolov2012statistical}
T.~Mikolov \emph{et~al.}, ``Statistical language models based on neural
  networks,'' \emph{Presentation at Google, Mountain View, 2nd April}, vol.~80,
  p.~26, 2012.

\end{thebibliography}
